\documentclass[aps,twocolumn,prl,floatfix,groupedaddress,nofootinbib,notitlepage,showpacs,superscriptaddress]{revtex4-1}

\usepackage{amsmath}
\usepackage{amsthm}
\usepackage{latexsym}
\usepackage{amssymb}
\usepackage{bm}
\usepackage{bbm}
\usepackage{appendix}
\usepackage{graphics,epstopdf}
\usepackage{physics}
\usepackage{color}
\usepackage[colorlinks=true,linkcolor=blue,citecolor=green,plainpages=false,pdfpagelabels]{hyperref}
\usepackage[normalem]{ulem}
\usepackage{amsmath}
\usepackage{appendix}
\usepackage{booktabs}
\usepackage{newlfont}
\usepackage{amsfonts}
\usepackage{amsthm}
\usepackage{times}
\usepackage{braket}
\usepackage{graphicx}
\usepackage{epsfig}
\usepackage{mathrsfs}
\usepackage[thinc]{esdiff}
\usepackage{caption}
\usepackage{subcaption}

\usepackage{times}

\DeclareOldFontCommand{\rm}{\normalfont\rmfamily}{\mathrm}

\newcommand{\off}[1]{}
\usepackage{bigints}
\usepackage{enumitem}

\usepackage{amsthm}
\usepackage{latexsym}
\usepackage{float}
\usepackage{appendix}
\usepackage{epstopdf}
\usepackage{physics}
\usepackage[colorlinks=true,linkcolor=blue,citecolor=green,plainpages=false,pdfpagelabels]{hyperref}
\usepackage[normalem]{ulem}
\usepackage{newlfont}
\usepackage{amsfonts}
\usepackage{epsfig}
\usepackage{mathrsfs}
\usepackage[thinc]{esdiff}

\DeclareOldFontCommand{\rm}{\normalfont\rmfamily}{\mathrm}
\usepackage{bigints}
\usepackage{braket}
\usepackage{caption}
\usepackage{subcaption}
\usepackage{setspace}
\usepackage{amsmath}
\usepackage{times,bm,bbm,bbold,graphicx,graphics,amssymb,amsmath,amsfonts,dsfont,hyperref,color}
\usepackage{mathrsfs,cancel}
\usepackage[utf8]{inputenc}
\usepackage{soul,xcolor}
\setstcolor{red}
\definecolor{mycolor}{rgb}{0.1, 0.1, 0.7}
\usepackage{dsfont}
\usepackage{url}
\usepackage{soul}
\usepackage[vlined,ruled]{algorithm2e}
\DeclareFontFamily{OT1}{pzc}{}
\DeclareFontShape{OT1}{pzc}{m}{it}%
{<-> s * [1.25] pzcmi7t}{}
\DeclareMathAlphabet{\mathpzc}{OT1}{pzc}%
{m}{it}
\hypersetup{
	plainpages=true,
	breaklinks=true,
	hypertexnames=false,
	pageanchor=true,
	colorlinks=true,
	linkcolor={mycolor},
	citecolor={mycolor},
	urlcolor={mycolor},
	anchorcolor={black}
}

\begin{document}

\title{Unified Framework for Direct and Complete Characterization of an Unknown Kraus Operator and Density Matrix Using a Single Input State}

\author{Sahil}
\email{sahil402b2@gmail.com}
\affiliation{Optics and Quantum Information Group, The Institute of Mathematical Sciences, CIT Campus, Taramani, Chennai 600 113, India}
\affiliation{Homi Bhabha National Institute, Training School Complex, Anushakti Nagar, Mumbai 400 085, India}

\author{Swarup Kumar Giri}
\email{physicsme1729@gmail.com}
\affiliation{School of Physical Sciences, National Institute of Science Education and Research, Jatni 752050, India }
\affiliation{Homi Bhabha National Institute, Training School Complex, Anushakti Nagar, Mumbai 400 085, India}

\author{Sohail}
\email{sohail.sohail@ttu.edu}
\affiliation{Department of Computer Science, Texas Tech University, Lubbock, TX 79409, USA}

\begin{abstract}
Characterization of quantum measurements and dynamical processes is typically performed using pure state preparations. However, in realistic experimental settings, the preparation of pure states is often infeasible due to noise and system constraints. In this work, we present a unified framework that enables the direct and complete characterization of an unknown Kraus operator using only a single input state. The same framework also supports the characterization of unknown observable, unitary operator, and density matrix. Remarkably, all these tasks are accomplished using a single input state, a set of projector-based unitary evolution operators, and the measurement of a single observable. Importantly, our approach imposes no constraints on the strength of the coupling between the system and a probe.
\end{abstract}

\maketitle

\section{Introduction}
Generalized quantum measurements are most effectively described using \textit{Kraus operators}, which capture both the post-measurement probability distributions and the resulting quantum state after measurement. At the core of quantum measurement theory are \textit{Positive Operator-Valued Measures} (POVMs), expressible via Kraus operators. These generalize ideal projective measurements to accommodate a wider class of measurement processes~\cite{Nielsen-Chuang-2010,Jacobs-2014}. In realistic scenarios---particularly when systems interact with their environments---POVMs offer a complete description of the measurable statistics accessible to observers. The standard technique for reconstructing these measurements is \textit{quantum process tomography}, which provides a full characterization of the POVM elements~\cite{Nielsen-Chuang-2010,Mohseni-Rezakhani-Lidar-2008}. While robust, this method is highly resource-intensive and inefficient when only partial information about specific POVM elements is needed.\par
To mitigate this inefficiency, recent work has explored the \textit{direct characterization of quantum measurements} (DCQM), which enables the selective determination of matrix elements of individual POVM elements with significantly reduced resource requirements~\cite{Xu-2021-DirectMeasurements,Kim-Kim-Lee-Han-Cho-2018}. However, existing DCQM techniques often rely on weak probe-system-environment interactions, sequential measurements, and auxiliary resources, thereby limiting their precision and scalability in high-dimensional settings~\cite{Kim-Kim-Lee-Han-Cho-2018}.\par
A deeper understanding of quantum measurements necessitates the identification of the \textit{Kraus operators} responsible for generating the observed statistics. This is nontrivial because a given POVM element can correspond to multiple distinct Kraus representations. For instance, consider the POVM $\left\{ \frac{1}{2}I, \frac{1}{8}(I + \sigma_Z), \frac{1}{8}(3I - \sigma_Z) \right\}$, where $\sigma_Z$ denotes the Pauli $Z$ operator. The element $E_0 = \frac{1}{2}I$ may be implemented using different Kraus operators, such as $A_0 = \frac{1}{\sqrt{2}}I$ or $\widetilde{A}_0 = \frac{1}{2}(\sigma_X + \sigma_Z)$ satisfying \(E_0 = A_0^\dagger A_0 = \widetilde{A}_0^\dagger \widetilde{A}_0\). This ambiguity highlights the need for methodologies capable of identifying both the POVM elements and their underlying Kraus structures. Moreover, standard quantum process tomography techniques and DCQMs typically rely on pure-state preparations, which are often unfeasible in practical scenarios due to unavoidable system--environment interactions that induce decoherence and result in mixed states.\par
In parallel, accurate characterization of quantum states---especially the off-diagonal elements of the density matrix---is crucial for revealing non-classical phenomena such as \textit{coherence} and \textit{entanglement}, which underpin numerous quantum technologies~\cite{Horodecki2009,Modi2010,Streltsov2017,Degen2017,Braun2018,Giovannetti2011,Vidrighin2014}. The conventional method for such characterization is \textit{quantum state tomography} (QST), which reconstructs the complete density matrix using an informationally complete set of measurements~\cite{James-2001,Thew-2002}. However, both the experimental and computational demands increase significantly with system size, making the QST technique impractical for large-scale systems.\par
To overcome these limitations, the \textit{direct characterization of the density matrix} (DCDM) has been proposed, enabling efficient extraction of specific matrix elements without requiring full state reconstruction~\cite{Lundeen‑Sutherland‑Patel‑Stewart‑Bamber‑2011,Lundeen-Bamber-2012,Thekkadath-2016,Vallone-Dequal-2016,Vallone-2018,Xu-Zhou-2024}. While DCDM reduces resource consumption, existing implementations typically rely on sequential coupling schemes, which become inefficient and error-prone in high-dimensional systems. Moreover, obtaining high-order correlations through sequential weak measurements amplifies statistical noise and reduces accuracy~\cite{Lundeen-Bamber-2012,Thekkadath-2016}. To enhance precision, strong-measurement-based strategies have been introduced~\cite{Vallone-Dequal-2016,Vallone-2018,Xu-Zhou-2024}, though these depends on the complex structures of the unitary operators acting on the system and a probe(s), and often demand additional assumptions or resources.\par
Here, we propose a \textit{unified framework} for the direct and complete characterization of an unknown Kraus operator, its associated POVM element, and a density matrix, using only a single input state. The framework further supports the characterization of an unknown observable and unitary operator. Remarkably, all these tasks are accomplished using a single input state, a set of projector-based unitary evolution operators, and the measurement of a single observable. Notably, our approach imposes no constraints on the coupling strength between the system and a probe. Unlike existing DCQMs and DCDMs, our framework avoids the need for complex unitary evolution, additional probes, or extra resources. We compare our method with existing DCQM and DCDM protocols in terms of the number of required unitary operations and measurement settings.
\section{Preliminaries: Kraus operators}
In this section, we discuss the origin and properties of Kraus operators. When a quantum system evolves in isolation, its dynamics are governed by a unitary operator. However, when the system interacts with another quantum system—typically referred to as the environment—the evolution is no longer unitary. We denote the system of interest by $S$ and the environment by $E$, and their Hilbert spaces by $\mathcal{H}_{\text{S}}$ and $\mathcal{H}_{\text{E}}$, respectively. Let us assume that the dimensions of the system and the environment are $d_{\text{S}}$ and $d_{\text{E}}$, respectively. \par
Suppose the system and the environment are initially in pure states $\ket{\psi_{\text{S}}}$ and $\ket{\xi_{\text{E}}}$, respectively. The joint evolution is described by a global unitary operator $U_{\text{SE}}$ acting on the composite Hilbert space. The combined state after evolution is then given by
\begin{equation}
\ket{\Psi_{\text{SE}}(t)} = U_{\text{SE}} \left( \ket{\psi_{\text{S}}} \otimes \ket{\xi_{\text{E}}} \right).
\label{KO-1}
\end{equation}
Because the state $\ket{\Psi_{\text{SE}}(t)}$ is generally entangled, a projective (selective) measurement on the environment using the operator $\Pi^k_{\text{E}} = \ketbra{k_{\text{E}}}{k_{\text{E}}}$ collapses the system-environment state according to the Born rule:
\begin{align}
\ket{\Psi_{\text{SE}}^{(k_{\text{E}})}(t)} &= \frac{1}{\sqrt{p_k}} (I \otimes \Pi^k_{\text{E}}) \ket{\Psi_{\text{SE}}(t)} \nonumber \\
&= \frac{1}{\sqrt{p_k}} A_k \ket{\psi_{\text{S}}} \otimes \ket{k_{\text{E}}},
\label{KO-2}
\end{align}
where $p_k = \bra{\Psi_{\text{SE}}(t)} (I \otimes \Pi^k_{\text{E}}) \ket{\Psi_{\text{SE}}(t)}$ is the probability of obtaining outcome $k_{\text{E}}$. The Kraus operator $A_k$ acting on the system is defined by
\begin{equation}
A_k = \bra{k_{\text{E}}} U_{\text{SE}} \ket{\xi_{\text{E}}}.
\label{KO-3}
\end{equation}
After measurement, the system is in the state $A_k \ket{\psi_{\text{S}}}/\sqrt{p_k}$, and the environment is in state $\ket{k_{\text{E}}}$. For a given Kraus operator $A_k$, the corresponding POVM element is defined as  
\begin{align}
E_k := A_k^{\dagger} A_k. \label{KO-3-1}
\end{align}
The probability $p_k$, introduced in Eq.~(\ref{KO-2}), can then be expressed in terms of $E_k$ as  
$p_k = \braket{\psi_{\text{S}}|E_k|\psi_{\text{S}}}$.\par
The operator $A_k$ encodes the effects of the global unitary $U_{\text{SE}}$, the initial state $\ket{\xi_{\text{E}}}$ of the environment, and the measurement outcome via $\Pi^k_{\text{E}}$. These operators are generally non-unitary but satisfy the completeness relation
\begin{equation*}
\sum_{k=0}^{d_{\text{E}}-1} A_k^{\dagger} A_k = I_{\text{S}},
\end{equation*}
which follows from the conservation of probability, $\sum_{k=0}^{d_{\text{E}}-1} p_k = 1$.\par
In the case of a non-selective measurement, where the measurement outcomes are not recorded, the post-measurement state of the system-environment is given by
\begin{align*}
\rho_{\text{SE}}^{\text{(non-sel)}} &= \sum_{k=0}^{d_{\text{E}}-1} p_k \ketbra{\Psi_{\text{SE}}^{(k_{\text{E}})}(t)}{\Psi_{\text{SE}}^{(k_{\text{E}})}(t)} \nonumber \\
&= \sum_{k=0}^{d_{\text{E}}-1} A_k \ketbra{\psi_{\text{S}}}{\psi_{\text{S}}} A_k^\dagger \otimes \ketbra{k_{\text{E}}}{k_{\text{E}}}.
\end{align*}
To obtain the reduced state of the system, we perform a partial trace over the environment:
\begin{align}
\mathcal{E}\left( \ketbra{\psi_{\text{S}}}{\psi_{\text{S}}} \right) = Tr_{\text{E}}\left[ \rho_{\text{SE}}^{\text{(non-sel)}} \right] = \sum_{k=0}^{d_{\text{E}}-1} A_k \ketbra{\psi_{\text{S}}}{\psi_{\text{S}}} A_k^\dagger.\label{KO-4}
\end{align}
Here, the map $\mathcal{E}$ describes a transformation of the system state, known as a quantum operation or quantum channel. Formally, it is a linear map
\begin{align*}
\mathcal{E} : \mathcal{L}(\mathcal{H}_{\text{S}}) \rightarrow \mathcal{L}(\mathcal{H}_{\text{S}}),
\end{align*}
where $\rho \in \mathcal{L}(\mathcal{H}_{\text{S}})$ is a density operator on the system Hilbert space.

The map $\mathcal{E}$ is known to satisfy the following properties~\cite{Nielsen-Chuang-2010,Lidar-2020}: ($i$) linearity, ($ii$) trace preservation, and ($iii$) complete positivity. Such maps are referred to as quantum channels in the context of quantum information theory and the theory of open quantum systems. For a detailed discussion, see Refs.~\cite{Nielsen-Chuang-2010,Jacobs-2014,Lidar-2020}. If the initial state of the system is a mixed state $\rho_{\text{S}}$, then Eq. (\ref{KO-4}) generalizes to: $\mathcal{E}(\rho_{\text{S}}) = \sum_{k=1}^{d_{\text{E}}} A_k \rho_{\text{S}} A_k^\dagger$.\par

\section{Unified framework}
Although a unified framework of this type was recently introduced in Ref. \cite{Sahil-Unified-2025}, we reconstruct it here to provide a more comprehensive understanding of our characterization technique.\par
In the context of measurement theory, defining Kraus operators typically requires an interaction between the system and its environment. In our approach, we introduce a probe that interacts with both the system and the environment, allowing for the direct characterization of individual Kraus operators. However, for the purpose of characterizing the density matrix, the system-environment interaction can be neglected, as it does not play an essential role. Let $\mathcal{H}_{\text{P}}$, $\mathcal{H}_{\text{S}}$, and $\mathcal{H}_{\text{E}}$ represent the Hilbert spaces of the probe, system, and environment, respectively. The dimensions of the system and environment are denoted by $d_{\text{S}}$ and $d_{\text{E}}$, while the probe is assumed to be a qubit, \emph{i.e.,} $\dim(\mathcal{H}_{\text{P}}) = 2$. The computational basis states of the probe are $\ket{0_{\text{P}}}$ and $\ket{1_{\text{P}}}$.\par
Let the probe, system, and environment initially be prepared in the product state $\rho(0) = \ketbra{\chi_{\text{P}}}{\chi_{\text{P}}} \otimes \rho_{\text{S}} \otimes \ketbra{\xi_{\text{E}}}{\xi_{\text{E}}}$. We consider the joint unitary evolution operator for the probe-system-environment as
\begin{align}
\!\!\!U_{\text{\scriptsize{PSE}}}=\ketbra{0_{\text{P}}}{0_{\text{P}}}\otimes U_{\text{S}} \otimes\! I_{\text{E}} + \ketbra{1_{\text{P}}}{1_{\text{P}}} \otimes U_{\text{SE}} \!\circ\!(\widetilde{U}_{\text{S}} \otimes I_{\text{E}}), \label{FW-1}
\end{align}
where $U_{\text{S}}$ and $\widetilde{U}_{\text{S}}$ are two different unitary operators acting on the system, and $U_{\text{SE}}$ represents an interaction unitary operator  between the system and the environment.
\begin{figure}[H]
\centering
\includegraphics[scale=0.46]{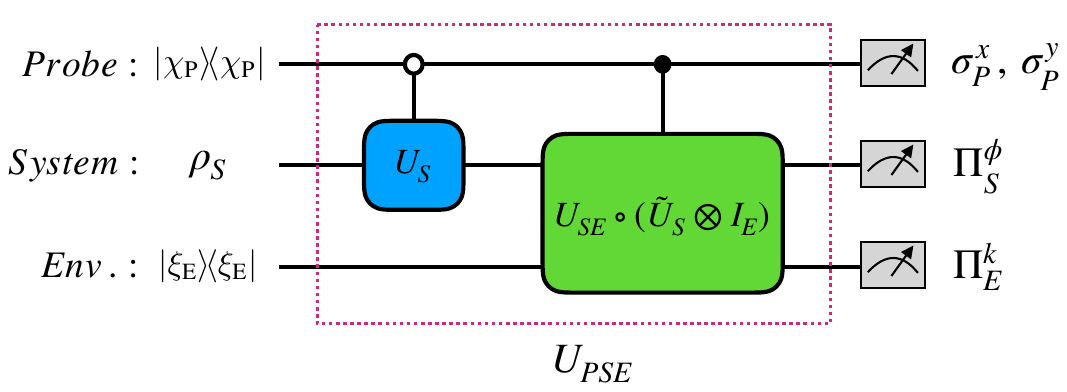}
\caption{Quantum circuit for implementation of the unitary operator $U_{\text{PSE}}$ given in Eq. \eqref{FW-1}.}
\label{FIG1}
\end{figure}
Equation~\eqref{FW-1} defines a controlled unitary evolution conditioned on the probe:
\begin{itemize}[leftmargin=*, itemsep=-2pt, topsep=3pt]
\item When the probe is in $\ket{0_{\text{P}}}$, the system evolves solely under the unitary operator $U_{\text{S}}$.\\
\item When the probe is in $\ket{1_{\text{P}}}$, the system and environment jointly evolve under the unitary operator $U_{\text{SE}} \circ (\widetilde{U}_{\text{S}} \otimes I_{\text{E}})$. This process is illustrated in Fig.~\ref{FIG1}.
\end{itemize}
After applying the global unitary $U_{\text{PSE}}$, the initial product state $\rho(0)$ evolves according to
\begin{align}
\rho(t) = U_{\text{PSE}}\, \rho(0)\, U_{\text{PSE}}^{\dagger}. \label{FW-2}
\end{align}
Subsequent measurements are performed on each subsystem: the Pauli operators $\sigma^x_{\text{P}}$ and $\sigma^y_{\text{P}}$ on the probe, projective measurements $\Pi^{\phi}_{\text{S}} = \ketbra{\phi_{\text{S}}}{\phi_{\text{S}}}$ on the system, and $\Pi^k_{\text{E}} = \ketbra{k_{\text{E}}}{k_{\text{E}}}$ on the environment. This yields the main result of this work:
\begin{align}
\braket{\phi_{\text{S}} | A_k \widetilde{U}_{\text{S}} \rho_{\text{S}} U_{\text{S}}^{\dagger} | \phi_{\text{S}}} 
= \frac{ \braket{ (\sigma^x_{\text{P}} + i \sigma^y_{\text{P}}) \otimes \Pi^{\phi}_{\text{S}} \otimes \Pi^{k}_{\text{E}} }_{\rho(t)} }{ \mathcal{N}_{\text{PSE}} }, \label{FW-3}
\end{align}
where $A_k = \braket{k_{\text{E}} | U_{\text{SE}} | \xi_{\text{E}}}$ is a Kraus operator defined in Eq. \eqref{KO-3}, $\mathcal{N}_{\text{PSE}} = 2 \braket{\chi_{\text{P}} | 0_{\text{P}}} \braket{1_{\text{P}} | \chi_{\text{P}}} \braket{\xi_{\text{E}} | k_{\text{E}}}$, and $\braket{ (\sigma^x_{\text{P}} + i \sigma^y_{\text{P}}) \otimes \Pi^{\phi}_{\text{S}} \otimes \Pi^{k}_{\text{E}} }_{\rho(t)}$ is the average value of the tripartite operator $(\sigma^x_{\text{P}} + i \sigma^y_{\text{P}}) \otimes \Pi^{\phi}_{\text{S}}\otimes \Pi^{k}_{\text{E}}$ \emph{w.r.t} the time evolved tripartite density operator $\rho(t)$ given by Eq. \eqref{FW-2}. 
A comprehensive derivation of Eq.~(\ref{FW-3}) can be found in the Supplemental Material of Ref.~\cite{Sahil-Unified-2025}. The relation is exact and does not rely on any form of approximation. In what follows, we demonstrate that Eq.~(\ref{FW-3}) provides a unified route to obtain the matrix elements of a Kraus operator, a unitary transformation, or an observable, all from a single mixed input state. Moreover, the same framework applies to reconstruct the matrix elements of an arbitrary system density operator. Importantly, all these quantities can be accessed within one experimental configuration, without requiring the weak-coupling limit.

\section{Direct and complete characterization of an unknown Kraus operator}
In this section, we show that only a single mixed state can be used to characterize a Kraus operator using the preparation-evolution-measurement technique described in the derivation of Eq. (\ref{FW-3}).\par
To obtain the $ij$-th element of the Kraus operator $A_k$, we set:
\begin{itemize}
\item $\widetilde{U}_{\text{S}}=e^{-i\theta\Pi_{\text{S}}^j}=I+(e^{-i\theta}-1)\Pi_{\text{S}}^j$, where $\theta$ is the system parameter and $\Pi_{\text{S}}^j=\ketbra{j}{j}$, 
\item $U_{\text{S}}=I_{\text{S}}$,
\item $\ket{\phi_{\text{S}}} = \ket{i}$
\end{itemize}
in Eq.~(\ref{FW-3}), which yields the $ij$-th matrix element:
\begin{align}
\braket{i|A_k|j}=&\left[\frac{\braket{ (\sigma^x_{\text{P}} + i \sigma^y_{\text{P}}) \otimes \Pi^{i}_{\text{S}} \otimes \Pi^{k}_{\text{E}} }_{\rho(t)}}{\mathcal{N}_{\text{PSE}}}-\braket{i|A_k\rho_{\text{S}}|i}\right]\nonumber\\
&\times\frac{1}{(e^{-i\theta}-1)\braket{j|\rho_{\text{S}}|i}}.\label{1DFCKO-1}
\end{align}
In this context, $\rho_{\text{S}}$ is taken to be such that $\braket{j|\rho_{\text{S}}|i}\neq 0$ for specified $i$ and $j$. We choose the states $\ket{\chi_{\text{P}}}$ and $\ket{\xi_{\text{E}}}$ such that $\braket{\chi_{\text{P}}|0_{\text{P}}} \neq 0$, $\braket{1_{\text{P}}|\chi_{\text{P}}} \neq 0$, and $\braket{\xi_{\text{E}}|k_{\text{E}}} \neq 0$, ensuring $\mathcal{N}_{\text{PSE}}\neq 0$. Unitary operators of the form $e^{-i\theta \Pi_{\text{S}}^j}$ have been widely used in methods for characterizing POVMs and density matrices~\cite{Xu-2021-DirectMeasurements, Vallone-2018, Xu-Zhou-2024}. The quantity $\braket{i|A_k\rho_{\text{S}}|i}$ is obtained from Eq. (\ref{FW-3}) by just setting $\widetilde{U}_{\text{S}}=I_{\text{S}}$, $U_{\text{S}}=I_{\text{S}}$, and $\ket{\phi_{\text{S}}}=\ket{i}$.\par
For the full characterization, we consider only one input state $\rho_{\text{S}}$ and $d_{\text{S}}+1$ number of unitary evolution operators $\{\{\widetilde{U}_{\text{S}}=e^{-i\theta\Pi_{\text{S}}^j}\}_{j=0}^{d_{\text{S}}-1},\widetilde{U}_{\text{S}}=I_{\text{S}}\}$, and for each unitary evolution operator $\widetilde{U}_{\text{S}}$, we have to measure an observable on the system whose eigenvectors are exactly \(\{\ket{\phi_{\text{S}}}\}=\{\ket{i}\}=\{\ket{0}, \ket{1}, \cdots, \ket{d_{\text{S}} - 1}\}\); see the construction and derivation of Eq.~(\ref{FW-3}). By doing so, we obtain the \(j\)-th column of the Kraus operator \(A_k\), and consequently, the entire Kraus operator \(A_k\). To ensure that $\braket{j|\rho_{\text{S}}|i} \neq 0$ for all $i,j$, we may initialize the system in a pure state of the form $\ket{\psi_{\text{S}}} = \sum_{i=0}^{d_{\text{S}} - 1} \alpha_i \ket{i}$ with all coefficients $\alpha_i \neq 0$. Passage through a generic noisy channel that transforms this state into a mixed state while preserving the condition $\braket{j|\rho_{\text{S}}|i} \neq 0$ for all $i,j$, thereby satisfying a key requirement of our protocol irrespective of the specific channel dynamics.\par
In practical scenarios, preparing a pure state is challenging due to unavoidable system-environment interactions that induce decoherence, ultimately resulting in a mixed state. Consequently, pure-state-based characterization schemes may either fail or yield significant errors in these settings; see for example \cite{Sahil-Unified-2025,Xu-2021-DirectMeasurements}. The method presented here overcomes this limitation by providing the matrix elements and complete characterization of a Kraus operator using a  mixed state only. It is straightforward to demonstrate that the error in estimating \( A_k \) using our method, as outlined in Eq.~\eqref{1DFCKO-1}, is approximately same to that in the method presented in Ref.~\cite{Sahil-Unified-2025}.

\section{Characterization of a POVM element}
Given a Kraus operator \( A_k \), the corresponding POVM element is defined as \( E_k := A_k^{\dagger}A_k \) [see Eq. \eqref{KO-3-1}]. The \( ij \)-th matrix element of \( E_k \) can be evaluated using
\begin{align}
\braket{i|E_k|j} = \sum_{l=0}^{d_{\text{S}}-1} \braket{l|A_k|i}^* \braket{l|A_k|j}. \label{DCPOVM-1}
\end{align}
To determine the sets \( \{ \braket{l|A_k|i} \}_{l=0}^{d_{\text{S}}-1} \) and \( \{ \braket{l|A_k|j} \}_{l=0}^{d_{\text{S}}-1} \) via Eq.~(\ref{1DFCKO-1}), we require an arbitrary input state $\rho_{\text{S}}$ such that $\braket{j|\rho_{\text{S}}|i}\neq 0$, two unitaries $\widetilde{U}_{\text{S}}=e^{-i\theta\Pi_i}$ and $\widetilde{U}_{\text{S}}=e^{-i\theta\Pi_{\text{S}}^j}$.  Then an observable is measured on the system whose eigenvectors are exactly \(\{ \ket{l} \}_{l=0}^{d_{\text{S}}-1} \), which remains unchanged across both the unitaries. In comparison, the method in Ref.~\cite{Xu-2021-DirectMeasurements} requires \( d_{\text{S}} \) distinct input states corresponding to the complete basis \( \{ \ket{s} \}_{s=0}^{d_{\text{S}} - 1} \), one unitary operator $e^{-i\theta\sigma_y\otimes\Pi_{\text{S}}^j}$ on the probe-system, where $\sigma_y$ is the Pauli-Y operator of the probe, and the measurement of the POVM element $E_k$ itself to access the same matrix element of $E_k$.\par
For full characterization of the POVM element \( E_k \), one may exploit the full matrix form of the corresponding Kraus operator \( A_k \) that directly gives \( E_k := A_k^{\dagger}A_k \).

\section{Characterization of an unknown unitary operator and observable}
Unitary operators play a crucial role in quantum information, including in the evolution of open quantum systems, the generation of entanglement, quantum measurements, and more \cite{Nielsen-Chuang-2010,Jacobs-2014,Horodecki2009,Lidar-2020,Breuer-book}. The characterization of such operators is therefore of significant interest. Similarly, observables, or projective measurements, are fundamental in quantum information theory. In many cases, it is of practical interest to determine which type of observables generate specific probability statistics or cause disturbances due to the strong or weak interaction between the probe and the system. In the following, we demonstrate how individual matrix elements of unknown unitary operators as well as observables can be accessed directly using a single mixed input state, when the interaction coefficient between the probe and the system is arbitrary.
\subsection{Unitary operator}
To obtain the $ij$-th element of an unknown unitary operator operator $U^1_{\text{S}}$, we set:
\begin{itemize}
\item $U_{\text{SE}}=I_{\text{SE}}\implies A_k=I_{\text{S}}$, also discard any involvement of the environment,
\item $\widetilde{U}_{\text{S}}=U^1_{\text{S}}\circ U^2_{\text{S}}$,
\item $U^2_{\text{S}}=e^{-i\theta\Pi_{\text{S}}^j}=I+(e^{-i\theta}-1)\Pi_{\text{S}}^j$, where $\theta$ is the system parameter and $\Pi_{\text{S}}^j=\ketbra{j}{j}$, 
\item $U_{\text{S}}=I_{\text{S}}$,
\item $\ket{\phi_{\text{S}}} = \ket{i}$
\end{itemize}
in Eq.~(\ref{FW-3}), which yields the $ij$-th matrix element:
\begin{align}
\braket{i|U^1_{\text{S}}|j}=&\left[\frac{\braket{ (\sigma^x_{\text{P}} + i \sigma^y_{\text{P}}) \otimes \Pi^{i}_{\text{S}}}_{\rho(t)}}{\mathcal{N}_{\text{PS}}}-\braket{i|U^1_{\text{S}}\rho_{\text{S}}|i}\right]\nonumber\\
&\times\frac{1}{(e^{-i\theta}-1)\braket{j|\rho_{\text{S}}|i}}.\label{UC-1}
\end{align}
Here, $\rho_{\text{S}}$ is taken to be such that $\braket{j|\rho_{\text{S}}|i}\neq 0$ for specified $i$ and $j$. We choose the state $\ket{\chi_{\text{P}}}$ such that $\braket{\chi_{\text{P}}|0_{\text{P}}} \neq 0$, $\braket{1_{\text{P}}|\chi_{\text{P}}} \neq 0$, ensuring $\mathcal{N}_{\text{PS}}\neq 0$. The quantity $\braket{i|A_k\rho_{\text{S}}|i}$ is obtained from Eq. (\ref{FW-3}) by just setting  $\widetilde{U}_{\text{S}}=U^1_{\text{S}}$, $A_k=I_{\text{S}}$, $U_{\text{S}}=I_{\text{S}}$, $\ket{\phi_{\text{S}}}=\ket{i}$, and finally discard any involvement of the environment.\par
For the full characterization, we consider only one input state $\rho_{\text{S}}$ and $d_{\text{S}}+1$ number of unitary evolution operators $\{\{U^2_{\text{S}}=e^{-i\theta\Pi_{\text{S}}^j}\}_{j=0}^{d_{\text{S}}-1}, U^2_{\text{S}}=I_{\text{S}}\}$, and for each unitary evolution operator $U^2_{\text{S}}$, we have to measure an observable on the system whose eigenvectors are exactly \(\{\ket{\phi_{\text{S}}}\}=\{\ket{i}\}=\{\ket{0}, \ket{1}, \cdots, \ket{d_{\text{S}} - 1}\}\); see the construction and derivation of Eq.~(\ref{FW-3}). By doing so, we obtain the \(j\)-th column of the unitary operator $U^1_{\text{S}}$, and consequently, the entire unitary operator $U^1_{\text{S}}$.
\subsection{Observable}
\off{From the relation $U^1_{\text{S}}=e^{-i\theta A}$, one can extract the matrix element of  the observable $A$ by expanding the exponential and discard the second and higher order terms: $\braket{i|A|j}\approx\frac{1}{i\theta}(\delta_{ij}-\braket{i|U^1_{\text{S}}|j})$. But in many practical scenarios, weak approximation \emph{i.e.,} $\theta\approx 0$ may not required or applicable, and for such cases, we can apply the following technique used in Ref. \cite{Sahil-Unified-2025}. \par
Let $U^1_{\text{S}}$ be the concatenation of two unitaries $e^{-i\theta_1A}$ and $e^{i\theta_2A}$ such that $U^1_{\text{S}}=e^{-i\theta_1A}e^{i\theta_2A}=e^{-i(\theta_1-\theta_2)A}$. Then we have:
\begin{align}
\braket{i|U^1_{\text{S}}|j}&=\delta_{ij}-i\delta\theta\braket{i|A|j}+\mathcal{O}(\delta\theta^2)\nonumber\\
&\approx\delta_{ij}-i\delta\theta\braket{i|A|j},\label{OC-1}
\end{align}
where $\delta\theta = \theta_1 - \theta_2$, and we assumed $\theta_1 \approx \theta_2$ but explicitly not $\theta_{1,2} \approx 0$; that is, we do not assume the interaction parameter to be weak. Under $\theta_1 \approx \theta_2$, second and higher-order terms in the expansion vanish. The error which occur due to the approximation in Eq. \eqref{OC-1} can be less erroneous by selecting $\theta_1$ and $\theta_2$ arbitrarily close. Even the accuracy of approximation considered in Eq. \eqref{UC-1} can be improved by cleverly retaining the terms up to second order; see Ref. \cite{Sahil-Unified-2025}.\par
Hence, using the $ij$-th matrix element of the Unitary operator $U^1_{\text{S}}=e^{-i\theta_1A}e^{i\theta_2A}$, we can obtain the $ij$-th matrix element of the observable $A$ using Eq. \eqref{OC-1}.}\par
From the relation \( U^1_{\text{S}} = e^{-i\theta A} \), one can approximate the matrix elements of the observable \( A \) by expanding the exponential and neglecting second and higher-order terms: $\braket{i|A|j} \approx \frac{1}{i\theta}(\delta_{ij} - \braket{i|U^1_{\text{S}}|j})$. However, in many practical scenarios, the weak interaction approximation (\( \theta \approx 0 \)) may not be required or experimentally feasible. In such cases, we propose the following technique applicable for arbitrary interaction.\par
Consider \( U^1_{\text{S}} \) as a concatenation of two unitaries: $U^1_{\text{S}} = e^{-i\theta_1 A} e^{i\theta_2 A} = e^{-i(\theta_1 - \theta_2) A}$. Then, expanding to first order in \( \delta\theta = \theta_1 - \theta_2 \), we obtain:
\begin{align}
 \braket{i|A|j}\approx \frac{1}{i\delta\theta}\left(\delta_{ij} - \braket{i|U^1_{\text{S}}|j}\right), \label{OC-1}
\end{align}
assuming \( \theta_1 \approx \theta_2 \) but not necessarily \( \theta_{1,2} \approx 0 \). This approximation avoids reliance on weak interactions, while higher-order terms vanish as \( \theta_1 \to \theta_2 \). The accuracy of Eq.~\eqref{OC-1} improves with smaller \( \delta\theta \), and can be further enhanced by systematically including second-order corrections, as discussed in Appendix A\ref{Appendix-A}. Hence, each matrix element \( \braket{i|A|j} \) of the observable can be estimated directly from the corresponding matrix element of the unitary operator \( U^1_{\text{S}} \) using Eq. \eqref{OC-1}.\par

\section{Characterization of an unknown density matrix}
To obtain the $ij$-th element of the density matrix $\rho_{\text{S}}$, we set:
\begin{itemize}
\item $U_{\text{SE}}=I_{\text{SE}}\implies A_k=I_{\text{S}}$, also discard any involvement of the environment,
\item $\widetilde{U}_{\text{S}}=I_{\text{S}}$,
\item $U_{\text{S}}=U^1_{\text{S}}\circ U^2_{\text{S}}$,
\item $U^2_{\text{S}}=e^{-i\theta\Pi_{\text{S}}^j}=I+(e^{-i\theta}-1)\Pi_{\text{S}}^j$, where $\theta$ is the system parameter and $\Pi_{\text{S}}^j=\ketbra{j}{j}$, 
\item $\ket{\phi_{\text{S}}} = \ket{i}$
\end{itemize}
in Eq.~(\ref{FW-3}), which yields the $ij$-th matrix element:
\begin{align}
\braket{i|\rho_{\text{S}}|j}=&\left[\frac{\braket{ (\sigma^x_{\text{P}} + i \sigma^y_{\text{P}}) \otimes \Pi^{i}_{\text{S}}}_{\rho(t)}}{\mathcal{N}_{\text{PS}}}-\braket{i|\rho_{\text{S}}{U^1_{\text{S}}}^{\dagger}|i}\right]\nonumber\\
&\times\frac{1}{(e^{i\theta}-1)\braket{j|{U^1_{\text{S}}}^{\dagger}|i}}.\label{CDM-1}
\end{align}
We take $U^1_{\text{S}}$ to be such that $\braket{j|{U^1_{\text{S}}}^{\dagger}|i}\neq 0$ for specified $i$ and $j$. We choose $\ket{\chi_{\text{P}}}$ such that $\braket{\chi_{\text{P}}|0_{\text{P}}} \neq 0$ and $\braket{1_{\text{P}}|\chi_{\text{P}}} \neq 0$, ensuring $\mathcal{N}_{\text{PS}}\neq 0$. The quantity $\braket{i|\rho_{\text{S}}{U^1_{\text{S}}}^{\dagger}|i}$ is obtained from Eq. (\ref{FW-3}) by just setting $A_k=I_{\text{S}}$, $\widetilde{U}_{\text{S}}=I_{\text{S}}$, $U_{\text{S}}=U^1_{\text{S}}$, and $\ket{\phi_{\text{S}}}=\ket{i}$.\par
For the full characterization, we have the unknown input state $\rho_{\text{S}}$ and $d_{\text{S}}+1$ number of unitary evolution operators $\{\{U^2_{\text{S}}=e^{-i\theta\Pi_{\text{S}}^j}\}_{j=0}^{d_{\text{S}}-1}, U^2_{\text{S}}=I_{\text{S}}\}$, and for each unitary evolution operator $U^2_{\text{S}}$, we have to measure an observable on the system whose eigenvectors are exactly \(\{\ket{\phi_{\text{S}}}\}=\{\ket{i}\}=\{\ket{0}, \ket{1}, \cdots, \ket{d_{\text{S}} - 1}\}\). By doing so, we obtain the \(j\)-th column of the density matrix $\rho_{\text{S}}$, and consequently, the entire density matrix $\rho_{\text{S}}$.\par
In Ref. \cite{Vallone-2018}, the authors demonstrated that to fully characterize an unknown density operator, their scheme requires $d_\text{S}$ distinct unitary operations: 
$\{ U_A^i = e^{-i\theta \Pi^i_{\text{S}} \otimes Y_A} \otimes I_B \}_{i=0}^{d_{\text{S}}-1}$,
along with an additional unitary operator $U_B = e^{-i\theta \Pi^{b^0}_{\text{S}} \otimes Y_B} \otimes I_A$ for each $i$, acting on a tripartite system consisting of the system (S) and two probes (A and B). Here, $\Pi_{\text{S}}^{b^0} = \ketbra{b^0_{\text{S}}}{b^0_{\text{S}}}$, and $\ket{b^0_{\text{S}}} = \frac{1}{\sqrt{d_{\text{S}}}} \sum_{i=0}^{d_{\text{S}}-1} \ket{i}$. Additionally, three joint probe measurements and the measurement of an observable on the system with eigenvectors $\{\ket{0}, \ket{1}, \cdots, \ket{d_{\text{S}} - 1}\}$ are required. In Ref. \cite{Xu-Zhou-2024}, it was shown that to obtain the $ij$-th matrix element of $\rho_{\text{S}}$, their scheme requires $d_\text{S}$ distinct unitary operations:
$\{ U_j = e^{-ig H_{i,j} (I_{\text{S}} - 2 \Pi_{\text{S}}^j) H_{i,j} \otimes \sigma_{\text{P}}^y} \}_{j=0}^{d_{\text{S}}-1}$, which act on the system (S) and a probe (P). Here, $H_{i,j}$ represents the Hadamard transformation on the system operator $I_{\text{S}} - 2 \Pi_{\text{S}}^j$. In addition, two probe measurements, $\sigma_{\text{P}}^x$ and $\sigma_{\text{P}}^y$, along with the measurement of an observable on the system, whose eigenvectors are $\{\ket{0}, \ket{1}, \cdots, \ket{d_{\text{S}} - 1}\}$, are required. Both of the methods described in Refs. \cite{Vallone-2018} and \cite{Xu-Zhou-2024} involve complex unitary evolution operators, in contrast to the method described in Eq. \eqref{CDM-1}. Notably, if the unitary evolution operators are Pauli X-gates, a complete characterization of an unknown density matrix can be achieved using only $\frac{d_{\text{S}}}{2} + 1$ Pauli X-gates for even dimensions $d_{\text{S}}$, or $\frac{d_{\text{S}} - 1}{2} + 1$ Pauli X-gates for odd dimensions $d_{\text{S}}$, as shown in Ref. \cite{Sahil-Unified-2025}.
 
\section{Discussion and Conclusion}
Knowledge of a POVM element alone is insufficient to determine the post-measurement state, as the latter is defined by the action of a specific Kraus operator on the initial state. Since a single POVM element may correspond to multiple distinct Kraus operators, identifying the correct one is essential for accurate quantum information processing. Previous works have concentrated on directly characterizing POVM elements rather than the associated Kraus operators. In this work, we have presented a unified framework that enables the direct and complete characterization of an unknown Kraus operator, as well as a density matrix. Our framework further supports the characterization of an unknown unitary operator and observable. All characterization tasks are accomplished using only a single input state, $d$ projector-based unitary evolution operators, and the measurement of a single observable. Crucially, our method does not rely on weak system--probe coupling, nor does it require detailed modeling of complex interactions. This stands in contrast to existing approaches, which often involve complex unitary evolution operators acting on one or more probes and the system.\par
Given that quantum measurements—\emph{i.e.}, POVMs—are central to both foundational studies and practical applications, their precise characterization is of fundamental importance. However, the preparation of pure input states, as required in existing DCQM schemes, is often infeasible in experimental settings due to noise and system constraints. Our framework overcomes this challenge by enabling the complete characterization of unknown Kraus operators, and hence POVMs, using only a single input state.
\vspace{3mm}
\\
\emph{\textbf{Acknowledgments.—}} SKG expresses his gratitude to Prof. Samyadeb Bhattacharya for facilitating his visit to the Center for Quantum Science and Technology, International Institute of Information Technology, Hyderabad, India, during which this work was completed.

\bibliography{bib}

\newpage
\onecolumngrid
\appendix

\section{Appendix-A}\label{Appendix-A}
A better approximation for determining matrix element $\braket{i|A|j}$ can be achieved instead of the first order approximation of $\delta\theta$ used to obtain Eq. \eqref{OC-1} in the following way.\par
By subtracting the $ij$-th elements of the two unitary operators $U^1_{\text{S}}=e^{-i(\theta_1-\theta_2)A}$ and ${U^1_{\text{S}}}^{\dagger}=e^{i(\theta_1-\theta_2)A}$, we have
\begin{align*}
\braket{i|{U^1_{\text{S}}}^{\dagger}|j}-\braket{i|U^1_{\text{S}}|j}&=2i\delta\theta \braket{i|A|j}+\mathcal{O}(\delta\theta^3)\\
&\approx 2i\delta\theta \braket{i|A|j},
\end{align*}
where the third and higher order terms are discarded, and thus we obtain
\begin{align}
\braket{i|A|j}=\frac{1}{2i\delta\theta}\left[\braket{i|{U^1_{\text{S}}}^{\dagger}|j}-\braket{i|U^1_{\text{S}}|j}\right].\tag{A1}\label{A1}
\end{align}
By substituting the $ij$-th elements of the unitary operators $U^1_{\text{S}}$ and ${U^1_{\text{S}}}^{\dagger}$ from Eq.~(\ref{UC-1}) into Eq.~(\ref{A1}), a more accurate estimate of the $ij$-th element of the observable $A$ is obtained.

\end{document}